\documentclass{PoS}
\usepackage{amsmath,amsthm,amssymb,amsfonts}

\title{On the dynamical determination of strange parton distributions}

\ShortTitle{On the dynamical determination of strange parton distributions}

\author{\speaker{P. Jimenez-Delgado}\thanks{Supported by the Swiss National Science Foundation (SNF) under contract 200020-126691.}\\University of Zurich\\E-mail: \email{pjimenez@physik.uzh.ch}}

\abstract{The dynamical parton distributions of the nucleon are generated radiatively from positive definite (valencelike) input distributions at an optimally chosen low resolution scale ($Q_0^2 < 1$ GeV$^2$). For the strange distribution in particular, it has been \emph{assumed} that vanishing strange input distributions at this low scale is an appropriate choice. By confronting predictions derived from our (GJR08) NLO dynamical parton distributions with recent neutrino dimuon production measurement from NuTeV we show that this is indeed the case, and that little improvement is achieved by using a more general ansatz. Nevertheless, the data induce an asymmetry in the strange sea which is found to be small and positive in agreement with previous results.}

\FullConference{XVIII International Workshop on Deep-Inelastic Scattering and Related Subjects\\April 19 -23, 2010\\Convitto della Calza, Firenze, Italy}

\begin{document}
The \emph{dynamical} parton distributions of the nucleon at $Q^2 \gtrsim 1$ GeV$^2$ are QCD radiatively generated from {\em valencelike}\footnote{Valencelike refers to $a_f\!>\!0$ for {\em all} input distributions $xf(x,Q_0^2)\propto x^{a_f}(1-x)^{b_f}$, i.e., not only the valence but also the sea and gluon input densities vanish at small $x$.} positive definite input distributions at an optimally determined low input scale $Q_0^2\!<\!1$ GeV$^2$. Therefore the \emph{steep} small-Bjorken-$x$ behavior of structure functions, and consequently of the gluon and sea distributions, appears within the dynamical (radiative) approach mainly as a consequence of QCD-dynamics at $x \lesssim 10^{-2}$ \cite{Gluck:1994uf}. Alternatively, in the common ``standard'' approach the input scale is fixed at some arbitrarily chosen $Q^2_0>1$ GeV$^2$, and the corresponding input distributions are less restricted; for example, the mentioned {\em steep} small-$x$ behavior has to be {\em fitted} here.

Following the radiative approach, the well-known LO/NLO GRV98 dynamical parton distribution functions of \cite{Gluck:1998xa} have been recently updated \cite{Gluck:2007ck}, and the analysis extended to the NNLO of perturbative QCD in \cite{JimenezDelgado:2008hf}. In addition, in \cite{Gluck:2007ck, JimenezDelgado:2008hf} a series of ``standard'' fits were produced in (for the rest) exactly the same conditions as their dynamical counterparts. This allows us to compare the features of both approaches and to test the the dependence in model assumptions. These analyses have been further augmented with appropriate uncertainty estimations and have shown \cite{Gluck:2007ck, JimenezDelgado:2008hf} that, as expected, the associated uncertainties encountered in the determination of the parton distributions turn out to be larger in the ``standard'' case, particularly in the small-$x$ region, than in the more restricted dynamical radiative approach where, moreover, the ``evolution distance'' (starting at $Q_0^2<1$ GeV$^2$) is sizably larger \cite{Gluck:2007ck, JimenezDelgado:2008hf}.

Since the data sets used in all these analyses are insensitive to the specific choice of the strange quark distributions, the strange densities of the \emph{dynamical} distributions in \cite{Gluck:1994uf, Gluck:1998xa, Gluck:2007ck, JimenezDelgado:2008hf} have been generated entirely radiatively starting from vanishing strange input distributions:
\begin{equation}
s(x,Q_0^2)=\bar{s}(x,Q_0^2)=0
\label{nullstrange}
\end{equation}
at the \emph{low} input scale. In the the ``standard'' case, where $Q^2_0 > 1$ GeV$^2$, the strange input distributions were chosen $s(x,Q_0^2)=\bar{s}(x,Q_0^2)= \tfrac{1}{4}\large(\bar{u}(x,Q_0^2)+\bar{d}(x,Q_0^2)\large)$, as is conventional \cite{Gluck:2007ck,JimenezDelgado:2008hf}. In order to investigate the plausibility of the \emph{assumptions} in Eq.(\ref{nullstrange}), we confront here predictions derived from dynamical distributions determined in this way with data which are particularly sensitive to the strangeness content of the nucleon. For this purpose we have chosen the latest and most precise measurements of neutrino dimuon production from $\nu_\mu$- and $\bar{\nu}_\mu$-iron deep inelastic scattering (DIS) interactions of NuTeV \cite{Mason:2007zz}.

For details on the calculation of the dimuon cross-section, including a direct comparison of our predictions with the data, and further neccesary references, we refer to the original paper \cite{JimenezDelgado:2010pc}. Here it suffices for our purposes to note that the results obtained with the GJR08 distributions are in good agreement with the data, e.g. we get $\chi^2=$ 65 for 90 data points (see \cite{JimenezDelgado:2010pc} for more details on this). This agreement demonstrates the compatibility of the data with the conditions of Eq.~(\ref{nullstrange}) and shows that in the dynamical case, where the NLO input distributions are parametrized at an optimally chosen low input scale $Q_0^2\!=\!0.5$ GeV$^2$ \cite{Gluck:2007ck}, the strange sea can be generated entirely radiatively starting from:
\begin{equation}
s^+(x,Q_0^2)\equiv s(x,Q_0^2)+\bar{s}(x,Q_0^2)=0
\label{splusnull}
\end{equation}

Turning now to the asymmetry in the strange nucleon sea, it is well known (see \cite{JimenezDelgado:2010pc} for references) that the small differences between neutrino and antineutrino data induce a small difference between the strange and antistrange parton distributions. In order to evaluate this asymmetry within our framework, we parametrize a new input distribution:
\begin{equation}
s^-(x,Q_0^2)\equiv s(x,Q_0^2)-\bar{s}(x,Q_0^2)= N x^a (1-x)^b (1-\tfrac{x}{x_0})
\label{sminus}
\end{equation}
where $Q_0^2\!=\!0.5$ GeV$^2$ is (fixed to) the input scale of our GJR08 NLO fit \cite{Gluck:2007ck} and the function is constrained by the quark-number sum rule $\int_0^1dx\; s^-\!(x,Q_0^2)=0$. Eqs.~(\ref{splusnull}) - (\ref{sminus}) imply that the strange input distributions, $s(x,Q_0^2)$ and $\bar{s}(x,Q_0^2)$, will in turn be negative (positive) at the input scale, by construction. This is not a problem as long as at perturbative scales, say for $Q^2\!>\!1$ GeV$^2$, both strange distributions $s(x,Q^2)$ and $\bar{s}(x,Q^2)$ become manifestly positive due to the QCD evolution, as is the case \cite{JimenezDelgado:2010pc}.

After introducing the asymmetry, the $\chi^2$ value improves to 60 for 90 data points, although the predictions from the strange-symmetric GJR08 distributions are rather similar and the differences lie within the 1$\sigma$ bands. Further, the (anti)neutrino data prefer (smaller)larger values, i.e. the data favor an increase of the $s$ distribution and a decrease of the $\bar{s}$, in other words, a \emph{positive} asymmetry in the relevant $0.01\lesssim x \lesssim 0.1$ region; for $x$ values larger than about $0.1$ no significant changes are appreciated \cite{JimenezDelgado:2010pc}.

\begin{figure}[t]
\centering
\includegraphics[width=0.7\textwidth]{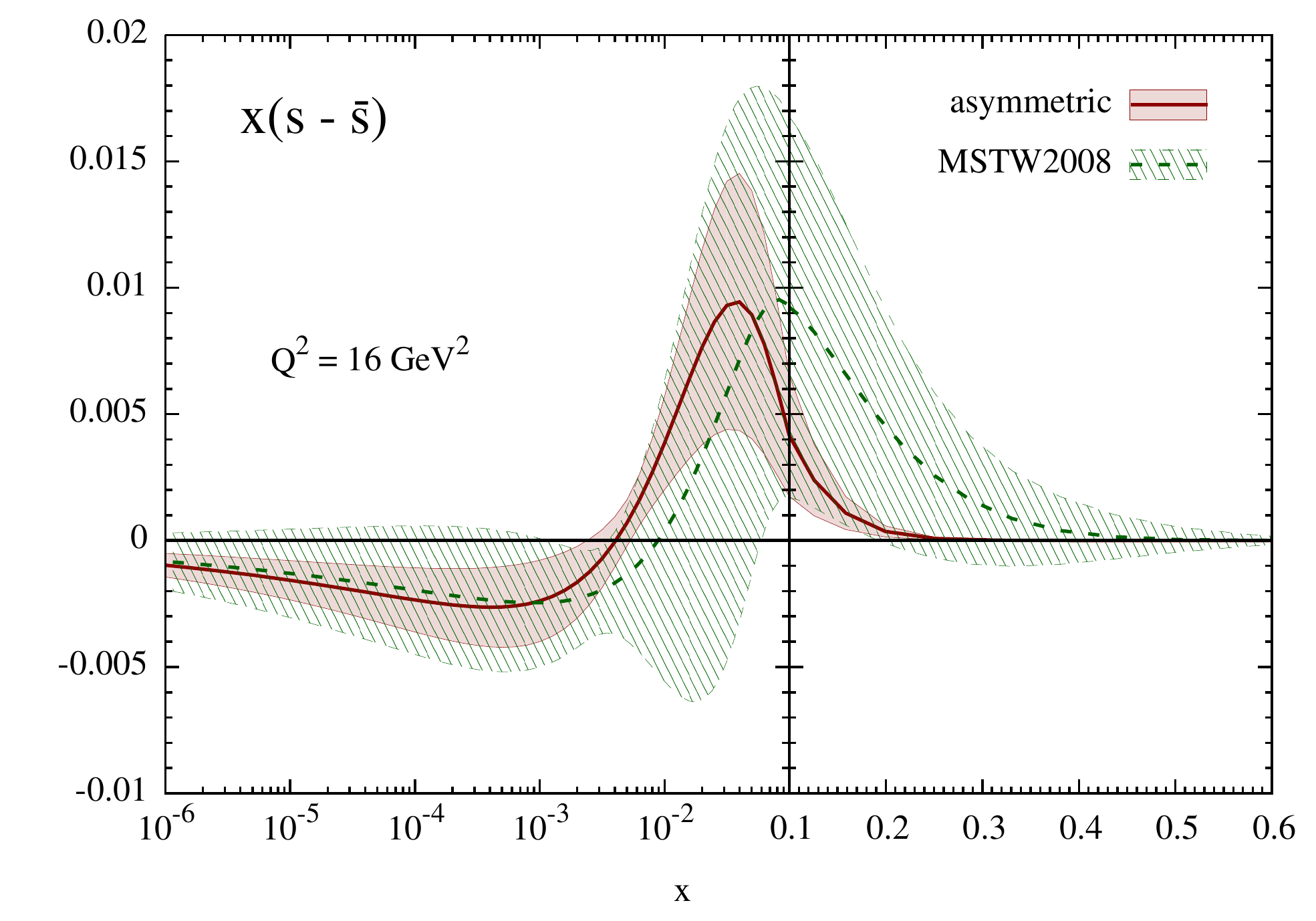}
\caption{Out result for the strangeness asymmetry in the nucleon at $Q^2\!=\!16$ GeV$^2$ appropriate for the NuTeV experiment (cf. Fig.~3 of \cite{Mason:2007zz}). The results of MSTW2008 \cite{Martin:2009iq} are also shown for comparison.\label{sasymmetry}}
\end{figure}

Our result for the strangeness asymmetry in the nucleon are shown in Fig.~\ref{sasymmetry} at $Q^2\!=\!16$ GeV$^2$ appropriate for the NuTeV experiment, and can be directly compared with Fig.~3 of \cite{Mason:2007zz}. Although due to the large errors both results are in general agreement, the peak of our asymmetry is lower and placed at a slightly smaller value of $x$. The results of MSTW2008 \cite{Martin:2009iq} are also shown in Fig.~\ref{sasymmetry} for comparison. They are rather similar in size to ours, despite the fact that in \cite{Martin:2009iq} older (and less precise) data have also been included and this tends to reduce the asymmetry \cite{Alekhin:2008mb}. Note, however, that our asymmetry is much more suppressed for large $x\!\gtrsim\!0.2$, where the data are in excellent agreement with our (strange-symmetric) GJR08 distributions.

The changes in the strange-asymmetric distributions as compared with the original GJR08 are rather small, e.g. at $Q^2\!=\!100$ GeV$^2$ they reach at most $5\%$ in the relevant $10^{-3} \!<\! x\! <\! 0.3$ region, and are comparable with the uncertainties in the distributions, which are of a few percent as well. This being the case, the original strange-symmetric GJR08 distributions should suffice for most applications, moreover since most observables depend essentially only on $s^+(x,Q^2)$. Furthermore, since we continue to generate the strange distributions radiatively starting from Eq.~(\ref{splusnull}), the increase in the uncertainties encountered in common ``standard'' fits, where $s^+(x,Q_0^2)$ has to be fitted, is avoided in the more constrained dynamical case, which uncertainties should be very similar to the ones of GJR08 in most cases.

The strangeness asymmetry is however relevant for applications especially sensitive to the strange content of the nucleon, as has been shown, for instance, in relation with the so-called the ``NuTeV anomaly'' (see, e.g. \cite{Gluck:2005xh} and references therein). As indication of the size and sign of the asymmetry it has become conventional to use the value of its second moment at the reference scale $Q^2\!=\!20$ GeV$^2$, we obtain:
\begin{equation}
S^- \equiv \int_0^1 dx\; x(s-\bar{s})=0.0008 \pm 0.0005
\end{equation}
which is of the right sign and size as to explain the ``anomaly'' and furthermore has, as expected, a relatively small error due to the dynamical assumptions. Previous determinations \cite{Mason:2007zz,  Martin:2009iq, Alekhin:2008mb} generally yield a larger value of about 0.0010 to 0.0020 and a typical uncertainty of about $100\%$ or even larger.

In conclusion, although in our global QCD fits \cite{Gluck:2007ck, JimenezDelgado:2008hf} no data with especial sensitivity to the strange content of the nucleon have been included, our determination of strange parton distributions, in particular Eq.~(\ref{splusnull}), is compatible with particularly sensitive data, e.g. those in \cite{Mason:2007zz}. Furthermore, these data induce an asymmetry in the strange sea which has been evaluated within our dynamical framework and found, in agreement with previous results, to be rather small and positive. This being the case, our original strange-symmetric distributions should suffice for most applications. The strangeness asymmetry may, however, be relevant for some especially sensitive applications; for these cases our results are available on request.

\end{document}